\documentclass[twocolumn,prd]{revtex4}

\usepackage{graphicx}% Include figure files
\usepackage{dcolumn}% Align table columns on decimal point
\usepackage{bm}% bold math

\newcommand{\be}{\begin{equation}}
\newcommand{\ee}{\end{equation}}
\newcommand{\ba}{\begin{eqnarray}}
\newcommand{\ea}{\end{eqnarray}}
\newcommand{\lsim}   {\mathrel{\mathop{\kern 0pt \rlap
  {\raise.2ex\hbox{$<$}}}
  \lower.9ex\hbox{\kern-.190em $\sim$}}}
\newcommand{\gsim}   {\mathrel{\mathop{\kern 0pt \rlap
  {\raise.2ex\hbox{$>$}}}
  \lower.9ex\hbox{\kern-.190em $\sim$}}}

\begin{document}
\title{Ultra-High Energy Neutrino Fluxes: New Constraints and Implications}
\author{Dmitry~V.~Semikoz$^{a,b}$,
G{\"u}nter Sigl$^c$}
\affiliation{
$^a$ Max-Planck-Institut f\"ur Physik (Werner-Heisenberg-Institut),\\
F\"ohringer Ring 6, 80805 M\"unchen, Germany\\
$^b$ Institute for Nuclear Research of the Academy
of Sciences of Russia,\\
Moscow, 117312, Russia\\
$^c$ GReCO, Institut d'Astrophysique de Paris, C.N.R.S., 98 bis boulevard
Arago, F-75014 Paris, France}

\begin{abstract}
We apply new upper limits on neutrino fluxes and
the diffuse extragalactic component of the GeV $\gamma-$ray flux
to various scenarios for ultra high energy cosmic rays and neutrinos.
As a result we find that extra-galactic top-down sources can not
contribute significantly to the observed flux of highest energy
cosmic rays. The Z-burst mechanism where ultra-high energy neutrinos produce
cosmic rays via interactions with relic neutrinos is practically
ruled out if cosmological limits on neutrino mass and clustering apply.
\end{abstract}

\maketitle

\section{Introduction}
High energy neutrino astrophysics is currently very active,
in particular experimentally~\cite{nu_tele}. Neutrino telescopes are
reaching sensitivities comparable to theoretical expectations for
neutrino fluxes based on their connection to primary cosmic rays
and secondary $\gamma-$rays~\cite{nu_review}. This is in particular
the case for upper limits from AMANDA II~\cite{amandaII}
in an energy range between $\simeq10^{14}\,$eV and $\simeq10^{18}\,$eV,
and from the RICE experiment~\cite{rice_new} above $10^{16}\,$eV.
The former aims to detect neutrinos by looking for showers and/or tracks
from charged leptons produced by charged current reactions of
neutrinos in ice, whereas the latter is searching for radio
pulses emitted by neutrino induced showers in south polar ice.
In addition, based on the non-observation of radio pulses from the
Earth's surface expected from neutrinos above $\sim10^{22}\,$eV,
the FORTE satellite has established upper limits on their fluxes
in this hitherto unexplored territory~\cite{forte}.

In the near future sensitivities will further improve by
next generation versions of these techniques. For AMANDA this
will be ICECUBE~\cite{icecube} at the South pole and possibly
a comparable kilometer scale neutrino telescope in the
Mediterranean~\cite{katz},
based on ANTARES~\cite{antares}, NEMO~\cite{nemo}, and NESTOR~\cite{nestor}.
Improved limits from the radio technique may come from
the Antarctic Impulsive Transient Antenna (ANITA) which is a
planned long duration balloon mission to detect radio waves
from showers induced by neutrinos in the antarctic ice~\cite{anita}.

Next generation experiments for ultra-high energy cosmic rays
(UHECR) above $\sim10^{19}\,$eV will also have considerable
sensitivity to neutrinos, typically from the near-horizontal
air-showers that are produced by them~\cite{auger_nu}.
These projects include the southern site of the Pierre Auger
Observatory~\cite{auger}, a combination of a charged
particle detector array with fluorescence telescopes for air showers
produced by cosmic rays above $\sim10^{19}\,$eV, and the telescope
array~\cite{ta}, which may serve as the optical component of
the northern Pierre Auger site. There are also plans
for space based observatories such as EUSO~\cite{euso} and OWL~\cite{owl}
of even bigger acceptance.

Finally, there are plans to construct telescopes to detect
fluorescence and \v{C}erenkov light from near-horizontal showers produced in
mountain targets by neutrinos in the intermediate window of
energies between $\sim10^{15}\,$eV and $\sim10^{19}\,$eV~\cite{fargion,mount}.
Acoustic detection of neutrino induced interactions is also
being considered~\cite{acoustic}.

In an earlier paper~\cite{kkss2} we reviewed fluxes in various scenarios
in the context of constraints from current cosmic ray data and upper
limits on $\gamma-$ray and neutrino fluxes. Besides the improved
neutrino flux limits from AMANDA II, RICE, and FORTE, a possibly lower
extragalactic contribution to the diffuse GeV $\gamma-$ray background observed
by the EGRET instrument on board the Compton $\gamma-$ray
observatory~\cite{egret,egret_new,kwl} has been pointed out recently.
An upper limit on the extragalactic
diffuse $\gamma-$ray flux constrains the total amount of electromagnetic
(EM) energy injected above $\sim10^{15}\,$eV which cascades down to below the
pair production threshold for photons on the cosmic microwave background
(CMB)~\cite{bere,bs}.
Since in any scenario involving pion production the EM energy fluence
is comparable to the neutrino energy fluence, a change in the constraint
on EM energy injection can also influence allowed neutrino fluxes.
Furthermore, future $\gamma-$ray detectors such as GLAST~\cite{glast}
will test whether the diffuse extragalactic GeV $\gamma-$ray background
is truly diffuse or partly consists of discrete sources that could
not be resolved by EGRET. This could further improve the cascade limit.

Motivated by these improved constraints and prospects, and by more
detailed information
available on neutrino sensitivities of future experiments, in the present
paper we reconsider flux predictions in scenarios currently often
discussed in the literature. As in Ref.~\cite{kkss2}, we apply
our recently combined propagation codes~\cite{code1,code2,photons,kkss}.
Sect.~II summarizes the numerical technique used in this paper.
In Sect.~III we discuss the cosmogenic neutrino flux, i.e. the
flux of neutrinos produced as secondaries of extragalactic cosmic
rays during propagation, and its dependence on various UHECR
source characteristics.
In Sect.~IV we review neutrino flux predictions in extragalactic top-down
scenarios where UHECRs are produced in decays of
super-massive particles continuously released from topological
defect relics from the early Universe.
If UHECRs are new hadrons, they have to be produced as secondaries
of accelerated protons which also gives rise to neutrinos.
Their fluxes in these scenarios are reviewed in Sect.~V.
Sect.~VI discusses primary neutrino fluxes required in scenarios where the
cosmic rays observed at the highest energies are produced as
secondaries from interactions with the
relic cosmological neutrino background, often called Z-burst
scenario. In Sect.~VII  we discuss neutrino fluxes from active galactic
nuclei (AGN) models.
Finally, in Sect.~VIII we conclude.

\section{Numerical Technique}
Our simulations use an implicit transport code
that evolve the spectra of nucleons, $\gamma-$rays, electrons,
electron-, muon-, and tau-neutrinos, and their anti-particles
along straight lines. Arbitrary injection spectra and
redshift distributions can be specified for the sources and
all relevant strong, electromagnetic, and weak interactions
have been implemented. For details see Refs.~\cite{code1,code2,kkss,kkss2}.
Our results apply to average large scale extragalactic magnetic
fields of the order $B\lesssim10^{-11}\,$G and a universal radio background
between the minimal values consistent with observations~\cite{Clark}
and moderately low theoretical estimates from radio source
counts~\cite{pb}.

For the neutrinos we assume for simplicity that all three flavors are
maximally mixed which for our purposes is an excellent
approximation~\cite{superK,mstv}
and thus have equal fluxes. For each flavor we sum fluxes of
particles and anti-particles.

In the present investigation we parameterize power law
injection spectra of either protons (for UHECR sources) or
neutrinos (for Z-burst models) per co-moving volume
in the following way:
\begin{eqnarray}
  \phi(E,z)&=&f(1+z)^m\,E^{-\alpha}\Theta(E_{\rm max}-E)\,
  \nonumber\\
  &&z_{\rm min}\leq z\leq z_{\rm max}\,,\label{para_inj}
\end{eqnarray}
where $f$ is the normalization that has to be fitted to the
data. The free parameters are the spectral index $\alpha$, the maximal
energy $E_{\rm max}$, the minimal and maximal
redshifts $z_{\rm min}$, $z_{\rm max}$, and the redshift
evolution index $m$. The resulting neutrino spectra depend insignificantly on
$z_{\rm min}$ in the range $0\leq z_{\rm min}\lesssim0.1$
where local effects could play a role, and thus we will set
$z_{\rm min}=0$ in the following.

To obtain the maximal neutrino fluxes for a given set of values for
all these parameters , we determine the maximal normalization $f$
in Eq.~(\ref{para_inj}) by demanding that both the accompanying
nucleon and $\gamma-$ray
fluxes are below the observed cosmic ray spectrum and
the diffuse $\gamma-$ray background observed by EGRET,
respectively.

\section{The Cosmogenic Neutrino Flux}
The flux of ``cosmogenic'' neutrinos is created by decaying charged
pions produced in interactions of primary nucleons of energy
above $\simeq5\times10^{19}\,$eV with CMB photons, the
Greisen-Zatsepin-Kuzmin (GZK) effect~\cite{gzk}.
This flux depends on the production
rate of the primary nucleons which we parameterize according to
Eq.~(\ref{para_inj}).

\subsection{Dependence on diffuse photon background}

\begin{figure}[ht]
\includegraphics[height=0.48\textwidth,clip=true,angle=270]{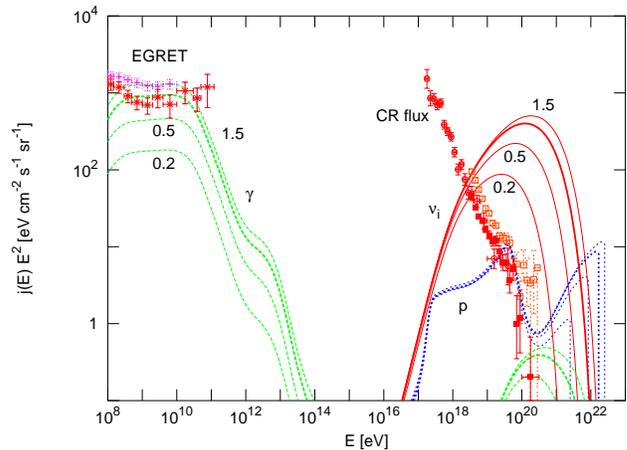}
\caption[...]{Dependence of the average cosmogenic neutrino flux
per flavor (labeled ``$\nu_i$'') on the contribution of the accompanying
photon flux (labeled ``$\gamma$'') to the old (upper error bars on
the left) and new (lower error bars on the
left) diffuse EGRET flux estimate. The proton primary parameters
in Eq.~(\ref{para_inj}) have been fixed to $m=3$, $z_{\rm max}=2$,
and $\alpha=1$, whereas $E_{\rm max}$ was varied. The UHECR proton flux
(labeled ``p'') is
normalized to the data of the AGASA~\cite{agasa} and HiRes~\cite{hires}
experiments. The numbers indicate the fraction of the new EGRET estimate
of the extragalactic diffuse $\gamma-$ray flux contributed by the
respective scenario, where the unlabeled thick curves correspond to 100\%.}
\label{F1}
\end{figure}

We first consider the dependence of the cosmogenic neutrino flux
on the contribution of the accompanying photons to the diffuse
$\gamma-$ray flux in the 100MeV - 100 GeV region. As an example,
we do this by fixing $m=3$, $z_{\rm max}=2$, and $\alpha=1$ in
Eq.~(\ref{para_inj}), while varying the maximal energy $E_{\rm max}$.
The relatively hard proton spectrum in this scenario could be produced,
for example, by acceleration in potential drops or reconnection~\cite{colgate}.
We normalize the resulting UHECR flux to the observations above
$10^{19}\,$eV and note that the discrepancy between the AGASA~\cite{agasa}
and HiRes~\cite{hires} fluxes above $\simeq10^{20}\,$eV has a
negligible influence on the predicted cosmogenic neutrino flux.
As Fig.~\ref{F1} shows, a decrease of the diffuse photon flux results from a
decrease of $E_{\rm max}$ in this case. The old EGRET flux
estimate~\cite{egret} corresponds to $E_{\rm max}=3\times 10^{22}\,$eV,
whereas the $\simeq50$\% smaller new EGRET flux~\cite{egret_new} corresponds
to $E_{\rm max}=2\times 10^{22}\,$eV. Most likely only a fraction of
the measured diffuse photon background is connected to GZK neutrinos. In the
present scenario 0.5 and 0.2 of the flux measured by EGRET corresponds
to $E_{\rm max}=10^{22}\,$eV and $E_{\rm max}=3\times 10^{21}\,$eV,
respectively. Note that the UHECR proton flux is the same in all cases,
except for the highest not yet observed energies where the flux can
be affected by the distance to the nearest sources.

The cosmogenic neutrino flux has recently been re-evaluated also
in Ref.~\cite{fkrt} where the EGRET constraint has not been
taken into account. The latter, however, eliminates a considerable
part of the higher fluxes considered there.

\subsection{Comparison with experimental limits and future sensitivities}

\begin{figure}[ht]
\includegraphics[height=0.48\textwidth,clip=true,angle=270]{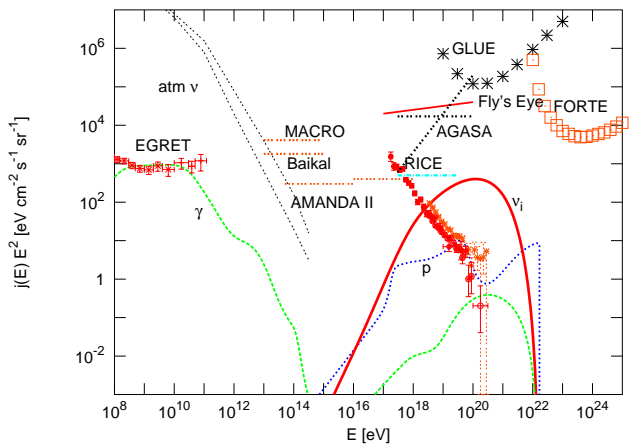}
\caption[...]{A scenario with maximal cosmogenic neutrino fluxes
per flavor as obtained by tuning the parameters of the proton
primaries in Eq.~(\ref{para_inj}) to $z_{\rm max}=2$,
$E_{\rm max}=2 \times 10^{22}\,$eV, $m=3$, $\alpha=1$. Also shown are
predicted and observed cosmic ray and $\gamma-$ray fluxes, the
atmospheric neutrino flux~\cite{lipari}, as well as existing upper
limits on the diffuse neutrino fluxes from MACRO~\cite{MACRO},
AMANDA II~\cite{amandaII}, BAIKAL~\cite{baikal_limit},
AGASA~\cite{agasa_nu}, the Fly's Eye~\cite{baltrusaitis} and
RICE~\cite{rice_new} experiments, and the limits obtained with the
Goldstone radio telescope (GLUE)~\cite{glue} and the FORTE
satellite~\cite{forte}, as indicated. The cosmic ray data are as in
Fig.~\ref{F1}, whereas only the new EGRET flux is shown to the left.}
\label{F2}
\end{figure}

\begin{figure}[ht]
\includegraphics[height=0.48\textwidth,clip=true,angle=270]{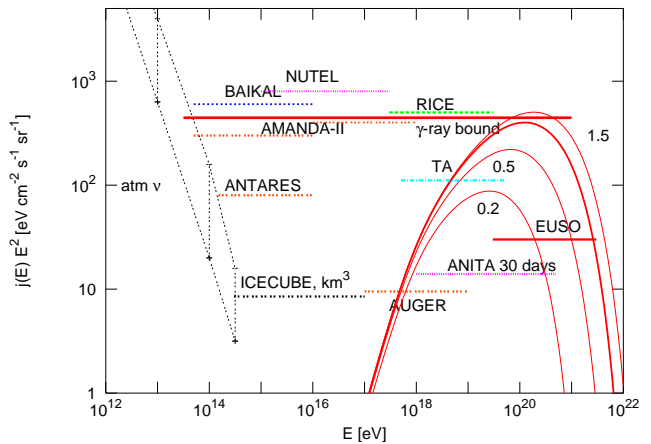}
\caption[...]{The cosmogenic neutrino flux per flavor shown in Fig.~\ref{F1}
in comparison with expected sensitivities of the currently being constructed
Pierre Auger project to  tau-neutrinos~\cite{auger_nu},
the planned projects telescope array (TA)~\cite{ta_nu}, the
fluorescence/\v{C}erenkov detector NUTEL~\cite{mount}, the
space based EUSO~\cite{euso_nu}, the water-based Baikal~\cite{baikal_limit}
and ANTARES~\cite{antares} (the NESTOR sensitivity for 1 tower would
be similar to AMANDA-II and for 7 towers similar to ANTARES~\cite{nestor}),
the ice-based AMANDA-II~\cite{amandaII} and ICECUBE~\cite{icecube}
(similar to the intended Mediterranean km$^3$ project~\cite{katz}), and
the radio detectors RICE~\cite{rice} and ANITA~\cite{anita}, as indicated.
All sensitivities except for ANITA and RICE refer to one year running
time. For comparison, the $\gamma-$ray bound derived from the EGRET GeV
$\gamma-$ray flux~\cite{egret_new} is also shown.}
\label{F3}
\end{figure}

\begin{figure}[ht]
\includegraphics[height=0.48\textwidth,clip=true,angle=270]{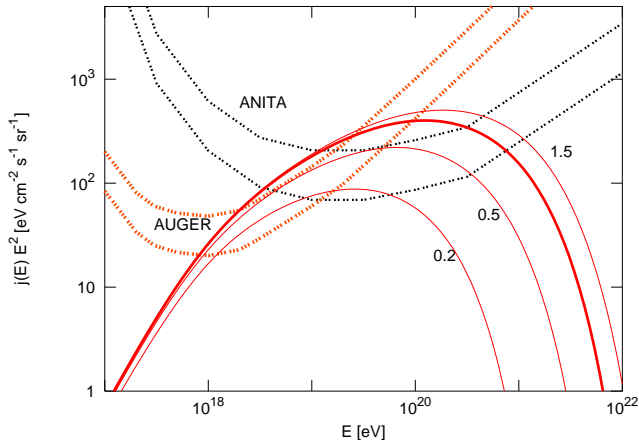}
\caption[...]{The cosmogenic neutrino flux average per flavor as shown
in Fig.~\ref{F1} in comparison with differential sensitivities expected
for 2006 for the Pierre Auger project~\cite{auger_nu} (for two years
of data taking, assuming strong (upper curve) and no (lower curve) deep
inelastic tau lepton scattering) and the ANITA project~\cite{anita} for 10
(upper curve) and 30 (lower curve) days of observation. The corresponding
number of events expected are listed in Tab.~\ref{tab}.}
\label{F4}
\end{figure}

The two major categories of experiments are based on detection in
water, ice or underground, typically sensitive below $\simeq10^{16}\,$eV,
and on air shower detection, usually sensitive at higher energies.
Existing neutrino flux upper limits come from the underground
MACRO experiment~\cite{MACRO}
at Gran Sasso, AMANDA II~\cite{amandaII} in the South Pole ice,
and the Lake BAIKAL neutrino telescope~\cite{baikal_limit} in the
first category, and the AGASA ground array~\cite{agasa_nu}, the
former fluorescence experiment Fly's Eye~\cite{baltrusaitis}, the
Radio Ice \v{C}erenkov Experiment RICE~\cite{rice_new} (there is also
a limit from the HiRes experiment~\cite{hires_nu}
which is between the RICE and AGASA limits), the Goldstone
Lunar Ultra-high energy neutrino experiment
GLUE~\cite{glue}, and the Fast On-orbit Recording of Transient Events
(FORTE) satellite~\cite{forte} in the second category. As an example,
an optimistic cosmogenic neutrino flux is compared with current neutrino flux
upper limits in Fig.~\ref{F2}. Future
experiments in the first category include NT200+ at Lake
Baikal~\cite{baikal_limit}, ANTARES~\cite{antares},
NESTOR in Greece~\cite{nestor}, as well as a possible common km$^3$
scale detector in the Mediterranean~\cite{nemo,katz}, and 
ICECUBE~\cite{icecube}, the
next-generation version of AMANDA at the South pole. The air shower based
category includes the Pierre Auger project~\cite{auger_nu}, the
telescope array~\cite{ta_nu}, the fluorescence/\v{C}erenkov detector
NUTEL~\cite{mount}, and the space based
EUSO~\cite{euso_nu} and  OWL~\cite{owl_nu} experiments.
The EUSO sensitivity estimate used here is based on deeply penetrating
atmospheric showers induced by electron or muon-neutrinos only~\cite{euso_nu}
and may thus be considerably better if tau neutrinos, \v{C}erenkov
events, and Earth skimming events are taken into account~\cite{skim},
for which there are no final estimates available yet.
The same applies to the OWL project~\cite{owl_nu}. The cosmogenic
neutrino flux models shown in Fig.~\ref{F1} are compared with future
sensitivities in Fig.~\ref{F3}.

The  fluxes shown in Figs.~\ref{F3} and~\ref{F4} are
considerably higher than the ones discussed in
Refs.~\cite{cosmogenic,stecker,pj,ydjs,ess}, and should be easily detectable
by at least some of these future instruments, as demonstrated by the
expected number of events listed in Tab.~\ref{tab}.

\begin{table}
\begin{center}
\begin{tabular}{|c|c|c|c|c|}
\hline
Experiment  &  $F_\nu(1.0)$ & $F_\nu(1.5)$ &  $F_\nu(0.5)$ &  $F_\nu (0.2)$ \\\hline
2 year Pierre Auger min&4.5&4.8&3.2&1.8\\ \hline
2 year Pierre Auger max&13.4&14.5&9.6&5.4\\ \hline
10 days ANITA&5.0&5.8&2.9&1.3\\ \hline
30 days ANITA&14.9&17.5&8.7&3.8\\
\hline
\end{tabular}
\end{center}
\caption{\label{tab}
The number of tau neutrino (Pierre Auger) and electron neutrino (ANITA) events
expected to be measured by 2006 for the four neutrino fluxes shown in
Figs.~\ref{F3} and~\ref{F4}. The Pierre Auger differential sensitivity
shown assumes two years of data taking until 2006.
The minimal (upper line in Fig.~\ref{F4}) and maximal (lower line in
Fig.~\ref{F4}) sensitivity depends on the assumptions made for the strength of
deep inelastic scattering of tau leptons. For ANITA the case for balloon
flights of 10 and 30 days are shown.}
\end{table}

We note that the non-observation of GZK neutrinos in 2006 will
significantly restrict the accompanying photon contribution to the
EGRET diffuse flux.

\section{Neutrino Fluxes in Top-Down Scenarios}

Historically, top-down (TD) scenarios were proposed as an alternative
to acceleration scenarios to explain the huge energies up to
$3\times10^{20}\,$eV observed in the cosmic ray spectrum~\cite{bhs}.
In these top-down scenarios UHECRs are the
decay products of some super-massive ``X'' particles of mass
$m_X\gg10^{20}\,$eV close to the grand unified scale, and have
energies all the way up to $\sim m_X$. Thus,
the massive X particles could be meta-stable relics of the early Universe
with lifetimes of the order the current age of the Universe
or could be released from topological defects
that were produced in the early Universe during
symmetry-breaking phase transitions predicted by in Grand Unified
Theories (GUTs). The X particles typically decay into leptons
and quarks. The quarks hadronize, producing jets of hadrons
which, together with the decay products of the unstable leptons,
result in a large cascade of energetic photons, neutrinos and light
leptons with a small fraction of protons and neutrons, some
of which contribute to the observed UHECR flux. The
resulting injection spectra have been calculated from QCD in
various approximations, see Ref.~\cite{bs} for a review and
Ref.~\cite{frag} for more recent work. In the present work
we will use the QCD spectra discussed in Ref.~\cite{lphdreview}
and shown in Fig.~11 of Ref.~\cite{kkss2}. For the purposes of
the current work this is not expected to make a significant difference
as compared to the more accurate fragmentation spectra~\cite{bss}.

\begin{figure}[ht]
\begin{center}
\includegraphics[angle=270,width=.48\textwidth,clip=true]{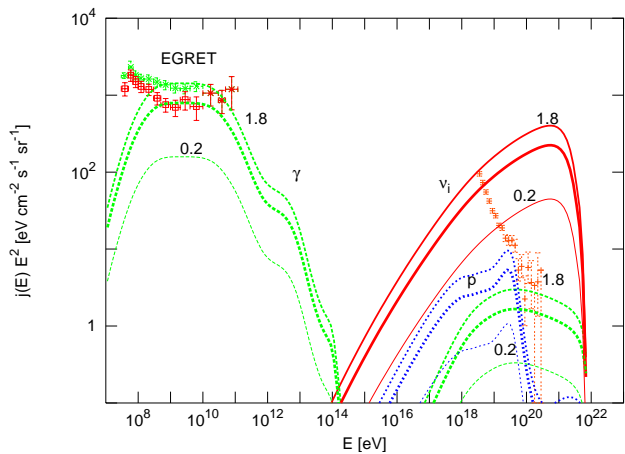}
\end{center}
\caption[...]{Flux predictions for a TD model characterized by $p=1$,
$m_X=2\times10^{13}\,$GeV. The contribution of photons and protons to
the UHECR flux decreases with decreasing fractional contribution to the
diffuse photon flux at EGRET energies which is denoted in numbers.
Even a TD contribution to the present estimate of the diffuse
EGRET flux as high as 100\% (unlabeled thick curves) is only marginally
consistent with the AGASA UHECR excess. The line key is the same as
in Fig.~\ref{F1}.
\label{F5}}
\end{figure}

For dimensional reasons the spatially averaged X particle
injection rate can only
depend on the mass scale $m_X$ and on cosmic time $t$ in the
combination
\begin{equation}
  \dot n_X(t)=\kappa m_X^p t^{-4+p}\,,\label{dotnx}
\end{equation}
where $\kappa$ and $p$ are dimensionless constants whose
value depend on the specific top-down scenario~\cite{bhs}.
Extragalactic topological defect sources usually predict
$p=1$, whereas decaying super-heavy dark matter (SHDM)~\cite{bkv97,kr97}
of lifetime much larger than the age of the Universe corresponds to
$p=2$~\cite{bs}. It has also been suggested that annihilation of SHDM
particles instead of their decay might contribute to the observed
UHECRs~\cite{BDK}.

In the SHDM scenario the observable UHECR flux will be dominated by
the decay or annihilation products of SHDM in the Galactic halo and
thus by sources at distances smaller than all relevant interaction
lengths. Composition and spectra will thus be directly given
by the injection spectra which are dominated by photons. This is most
likely inconsistent with upper limits on the ultra-high energy (UHE)
photon fraction above $10^{19}\,$eV~\cite{photon_limit}. However, requiring
that this scenario explains only UHECRs above $4\times10^{19}\,$eV
allows to avoid this problem~\cite{ks2003}.

A more severe  problem of the SHDM scenario is the spatial anisotropy
of the expected signal predicted due to the non-central position of the Sun in
our Galaxy \cite{dt1998}. In a recent paper~\cite{ks2003} it was shown that the
non-observation of anisotropy in the data of the SUGAR experiment
excludes any SHDM scenario at the $5\sigma$ level
assuming that all events above $4\times10^{19}\,$eV are from SHDM sources.
For the extreme case where
SHDM is responsible only for UHECRs above $6\times10^{19}\,$eV,
the annihilation scenario is disfavored at least at 99\% CL by the SUGAR
data, while decaying SHDM still has a probability of
$\sim 10\%$ to be consistent with the SUGAR data.

The SHDM scenario is therefore disfavored by present experimental data
and will be finally tested by the Pierre Auger experiment in the near future.
We will therefore here focus on topological defect models with $p=1$.

\begin{figure}[ht]
\begin{center}
\includegraphics[angle=270,width=.48\textwidth,clip=true]{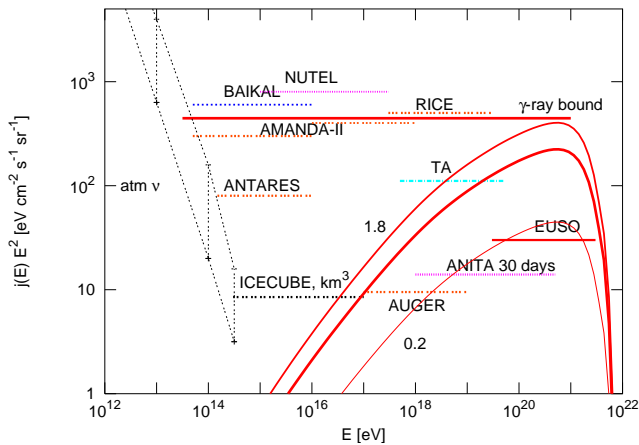}
\end{center}
\caption[...]{Neutrino fluxes per flavor predicted by the three normalizations
of the TD model of Fig.~\ref{F5} compared to future experimental
sensitivities. The line key is as in Fig.~\ref{F3}.
\label{F6}}
\end{figure}

Fig.~\ref{F5} shows the results for $m_X=2\times10^{13}\,$GeV,
with $B=10^{-12}\,$G, and the moderately low theoretical estimate
from Ref.~\cite{pb}. These parameters lead to optimistic neutrino fluxes
for the maximal normalization consistent with all data. For
detailed earlier discussions of extragalactic top-down fluxes
see Refs.~\cite{slsc,slby}.

Fig.~\ref{F5} shows that already the improved upper limit on the
true diffuse photon background by EGRET implies too small a
UHECR flux compared to the AGASA excess at energies $E\gtrsim10^{20}\,$eV.
The parameters used in the figure represent the ``best fit point'' for
this model; in particular for all other masses $m_X$ the disagreement is
more severe. The new EGRET upper limit thus strongly disfavors extragalactic
top-down scenarios. In addition, independently of this problem of
overproduction of GeV $\gamma-$rays, a non-observation of TD neutrinos by
2006 will rule out the possibility that extragalactic top-down mechanisms
significantly contribute to the UHECR flux, as can be seen from comparing
Fig.~\ref{F6} with Fig.~\ref{F5}.

\section{Neutrino flux in scenarios with new hadrons as UHECR primaries}

\begin{figure}[ht]
\begin{center}
\includegraphics[angle=270,width=.48\textwidth,clip=true]{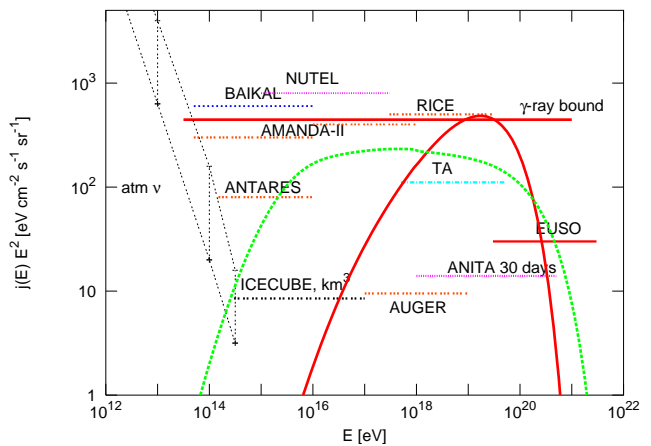}
\end{center}
\caption[...]{Neutrino fluxes per flavor predicted by scenarios where
UHECRs are explained as new hadrons produced as secondaries of accelerated
protons with $m=3$, $z_{\rm max}=2$, compared to expected
experimental sensitivities. The solid line is for a flux of primary
protons peaked at $E=10^{21}\,$eV and the dashed line is for
a primary proton flux $\propto E^{-2}$ up to $E_{\rm max}=10^{22}$ eV. The
line key is as in Fig.~\ref{F3}.
\label{F7}}
\end{figure}

Supersymmetric (SUSY) models with a strongly interacting particle
as lightest supersymmetric particle (LSP) or next-to-lightest SUSY
particle (NLSP) are very
interesting for explaining the AGASA excess above the GZK cutoff.
Hadrons containing a gluino were first suggested by Farrar as
UHECR primaries~\cite{Fa96,farrar}. This model with a light gluino
together with a light photino as cold dark matter candidate is meanwhile
excluded~\cite{exp_g1,exp_g2}.
However, more general models with a light gluino or a light sbottom
quark are still viable.

In a recent paper~\cite{kst2003} a model-independent, purely
phenomenological approach was developed. Since the observed extensive air
showers (EAS) are consistent with simulated EAS initiated by protons,
any new primary proposed to solve the GZK puzzle has to produce EAS
similar to those of protons. Experimentally still allowed
are photons as UHECR primaries: at 90\% CL, $\sim30\%$ of UHECRs
above $\sim10^{19}\,$eV can be photons~\cite{photon_limit}. However, the
simplest possibility consistent with EAS observations
is to require that the new primary is strongly interacting.
The requirements of efficient production in astrophysical accelerators
as well as proton-like EAS in the atmosphere ask for a
light hadron, $\lesssim3\,$GeV, while shifting the GZK cutoff to
higher energies results in a lower bound for its mass,
$\gtrsim1.5\,$GeV~\cite{Berezinsky:2001fy}. From these requirements
general conditions on the interactions of new UHE primaries were derived.
The production of new hadrons in astrophysical objects was investigated
in Ref.~\cite{kst2003}. It was found that proton-proton
collisions in astrophysical accelerators cannot produce sufficiently high
fluxes of new primaries without contradicting existing measurements of
photon~\cite{egret} and neutrino fluxes~\cite{amandaII,rice_new,forte}.
In contrast, for a light shadron with mass $\lesssim3\,$GeV and
the astrophysically more realistic case of UHE proton collisions on
optical/infrared background photons there is no contradiction with existing
limits. Also, the required initial proton energy is not too extreme,
$E\lesssim10^{21}\,$eV, which may be achieved by astrophysical acceleration
mechanisms. The only essential condition for the sources is that
they should be optically thick for protons in order to produce these
new hadrons. This condition applies to all models with new particles
produced by protons.

One of the important features of scenarios with new hadrons, and of
any model in which the production cross section $\sigma_{p\gamma\to S}$
of a new particle $S$ is much smaller than the total proton-photon cross
section $\sigma_{p\gamma}$, is the high flux predicted for secondary
high-energy neutrinos. This neutrino flux is approximately
$F_{\rm CR}\sigma_{p\gamma}/\sigma_{p\gamma\to S}$ in terms of
the maximal contribution of $S$ particles to the observed cosmic ray flux,
$F_{\rm CR}\simeq(E/10^{20}\,{\rm eV})^{-2}\,$~eV/(cm$^2$ s sr).

Fig.~\ref{F7} shows that for a primary proton flux $\propto E^{-2}$,
this model can be restricted already with 3 years of AMANDA-II data.
A non-observation of neutrinos by 2006 will make it impossible to
render the production cross section of the new hadrons consistent with
existing limits in these scenarios.

\section{The Z-Burst Scenario}
\begin{figure}[ht]
\begin{center}
\includegraphics[angle=270,width=.48\textwidth,clip=true]{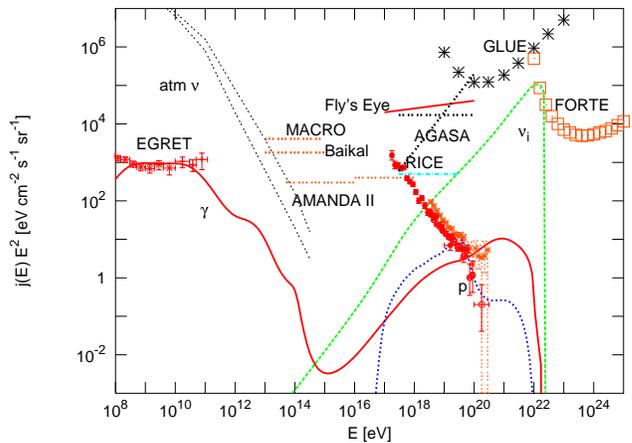}
\end{center}
\caption[...]{Flux predictions for a Z-burst model averaged over flavors and
characterized by the injection parameters $z_{\rm min}=0$, $z_{\rm max}=3$,
$\alpha=1$, $m=0$, $E_{\rm max}=3\times10^{22}\,$eV in
Eq.~(\ref{para_inj}) for neutrino primaries.
The sources are assumed to be exclusive neutrino emitters. All
neutrino masses were assumed equal with $m_\nu=0.33$~eV and we again
assumed maximal mixing between all flavors. The line key is as
in Fig.~\ref{F2}.
\label{F8}}
\end{figure}

In the Z-burst scenario UHECRs are produced by Z-bosons
decaying within the distance relevant for the GZK effect. These
Z-bosons are in turn produced by UHE neutrinos interacting with
the relic neutrino background~\cite{zburst1}. If the relic neutrinos
have a mass $m_\nu$, Z-bosons can be resonantly produced by UHE
neutrinos of energy
$E_\nu\simeq M_Z^2/(2m_\nu)\simeq4.2\times10^{21}\,{\rm eV}\,({\rm eV}/m_\nu)$.
The required neutrino
beams could be produced as secondaries of protons accelerated
in high-redshift sources. The fluxes predicted in these scenarios
have recently been discussed in detail in Refs.~\cite{fkr,kkss}.
In Fig.~\ref{F8} we show an optimistic example taken from Ref.~\cite{kkss}.
As in Refs.~\cite{fkr,kkss}
no local neutrino over-density was assumed. The sources
are assumed to not emit any $\gamma-$rays, otherwise the Z-burst
model with acceleration sources overproduces the diffuse GeV $\gamma-$ray
background~\cite{kkss}. We note that no known
astrophysical accelerator exists that meets the requirements
of the Z-burst model~\cite{kkss,gtt2003}.

However, a combination of new constraints allows to rule out
the Z-burst mechanism even for pure neutrino emitting sources:
A combination of cosmological data including the WMAP experiment limit
the sum of the masses of active neutrinos to $\lesssim1\,$eV~\cite{hannestad}.
Solar and atmospheric neutrino oscillations indicate that individual
neutrino masses are nearly degenerate on this scale~\cite{mstv}, and thus
the neutrino mass per flavor must satisfy $m_\nu\lesssim0.33\,$eV.
However, for such masses phase space constraints limit the possible
over-density of neutrinos in our Local Group of galaxies to
$\lesssim10$ on a length scale of $\sim1\,$Mpc~\cite{sm}. Since
this is considerably smaller than the relevant UHECR loss lengths,
neutrino clustering will not significantly reduce the necessary UHE neutrino
flux compared to the case of no clustering.
For the maximal possible value of the neutrino
mass $m_\nu \simeq 0.33\,$eV the neutrino flux required for the Z-burst
model is only in marginal conflict with the
FORTE upper limit~\cite{forte}, as shown in Fig.~\ref{F8}. For all other
cases the conflict is considerably more severe. Also note that
this argument does not depend on the shape of the low energy
tail of the primary neutrino spectrum which could thus be even
monoenergetic, as could occur in exlusive tree level decays
of superheavy particles into neutrinos~\cite{gk}. However, in addition
this possibility has been ruled out by overproduction of GeV
$\gamma-$rays due to loop effects in these particle decays~\cite{bko}.

As was discussed in Ref.~\cite{kkss}, the Z burst scenario involving
normal astrophysical sources producing neutrinos and photons by pion
production within the source were already ruled out by the former
EGRET limit.

\section{Neutrino fluxes in scenarios involving Active Galactic Nuclei}

\begin{figure}[ht]
\begin{center}
\includegraphics[angle=270,width=.48\textwidth,clip=true]{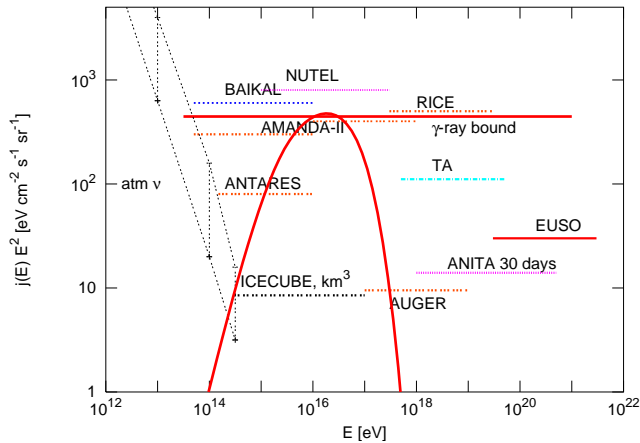}
\end{center}
\caption[...]{Neutrino flux per flavor predicted for the AGN model from
Ref.~\cite{neronov} for a uniform distribution of blazars (no redshift
evolution). The position of the peak is governed by the initial proton
distribution. The normalization is determined by the amplitude of the
accompanying $\gamma-$ray flux to the new diffuse EGRET flux estimate.
The line key is as in Fig.~\ref{F3}.
\label{F9}}
\end{figure}

Active Galactic Nuclei (AGN) are very promising sites of particle
acceleration. Though there is no direct evidence of proton acceleration
in these objects, according to the Hillas condition $E_{\rm max}\sim q B R$
for the maximal energy, where $q$ is the charge, $B$ is the magnetic field
and $R$ is the linear size of the acceleration region, AGN cores,
jets or hot spots can be sites for acceleration of UHECRs
up to the highest energies $E\sim 10^{19}-10^{21}$ eV. Once accelerated,
protons can escape the AGN freely or can lose part of their energy
in interactions with background protons and photons.
High energy neutrinos can be produced in AGN via pion production
by accelerated protons.

The total power of a given object in neutrinos can
be related to the total power in $\gamma-$rays of MeV-GeV energies.
However, such a connection is not straightforward because
the observed high energy $\gamma-$rays can be produced by several mechanisms
not involving neutrino production, for example proton synchrotron radiation
in magnetic fields and inverse Compton scattering of MeV-GeV electrons.
Moreover, the spectrum of neutrinos from AGN are even more difficult
to predict than the spectrum of cosmogenic neutrinos. The reason is
that besides the unknown model-dependent spectrum of primary protons,
the spectrum of background photons is also not known and in general
is model dependent. Also, AGN are divided into subclasses with different
properties of the observed photon spectrum. Many of those properties do
not directly relate to the possible neutrino spectrum. This means that
it is very difficult to predict the space distribution of those AGN
which contribute to the neutrino flux from the distribution of AGN
subclasses. At least these distributions can be very model dependent.

Due to the above complications we suggest here a phenomenological approach
to the prediction of neutrino fluxes. Within this
approach the neutrino flux in most of AGN models can be approximately
characterized by three parameters, namely the amplitude, width and position
of the peak (plateau) in the differential spectrum. The position of the
peak (plateau) is related to the spectrum of background
photons. The combination of amplitude and width defines the total power
in neutrinos which can be related to the total power in MeV-GeV photons
produced in the same pion production reactions. Experimental bounds on
neutrino fluxes can be converted to constraints on model parameters.

A similar approach can be used for most existing models which predict
neutrino fluxes from AGN, see for example Ref.~\cite{Stecker:1995th}.
The only difference would be the connection of the phenomenological parameters
of the observed neutrino spectrum to the physical AGN parameters in the
given model.

As an example, we will use the model of $\gamma-$ray powered jets of
Ref.~\cite{neronov}. In this model the high energy $\gamma$-rays are produced
by accelerated protons interacting with the ambient photon fields
(supplied, for example, by the accretion disk around the massive black hole)
through photo-meson processes. At the same time those protons produce neutrinos
which are emitted in the direction of the jet. Therefore, this model
predicts a high neutrino flux comparable in power with the
$\gamma$-ray flux. The detailed numerical simulations of proton acceleration
in the central engine of the AGN~\cite{ns2002ac} show that
the collimated jet of almost mono-energetic high energy protons
(linear accelerator)
can be created in the electro-magnetic field around the black hole
and the energy of those protons can be converted into photons and neutrinos,
while protons can be captured inside the source.
The nucleon flux leaving the AGN is well below the observed cosmic ray
flux in this scenario. Furthermore, since all nucleons leaving the
source are well below the GZK cutoff, there is no cosmogenic contribution
to the neutrino flux from these sources.

Fig.~\ref{F9} shows a typical prediction for the diffuse neutrino
flux in this model. The neutrino flux is maximized in such a way that
the accompanying photon flux saturates the new EGRET bound. As seen
from Fig.~\ref{F9}, already now AMANDA-II and RICE data start to
restrict AGN models. Three years of AMANDA-II data will significantly
restrict the parameter space of those models. This will also restrict
the contribution of $\pi^0$ production in AGN to the extragalactic
diffuse $\gamma-$ray flux at EGRET energies.

In the AGN  model discussed above, blazars would be seen by neutrino
telescopes as point-like sources with neutrino fluxes which are smaller or
of the same order as the photon flux emitted by these same sources and
which are detectable by $\gamma$-ray telescopes. The most probable
sources were discussed in Ref.~\cite{ns2002}.

\section{Conclusions}
Based on our transport code we reconsidered neutrino flux predictions
and especially their maxima consistent with all current data on
cosmic rays and updated upper limits on neutrino fluxes and the
diffuse extragalactic GeV $\gamma-$ray background.
We discussed predictions for fluxes of cosmogenic neutrinos produced through
pion production of UHECRs during propagation, and for fluxes produced by
AGN. We showed that extragalactic top-down scenarios can not
contribute significantly to the observed ultra-high energy cosmic
ray flux if standard evolution histories and injection spectra
are used. The Z-burst mechanism where ultra-high energy neutrinos produce
cosmic rays via interactions with relic neutrinos is
ruled out except if cosmological neutrino mass limits are
invalid and/or if relic neutrinos cluster more strongly than
expected based on standard phase space principles. Only the case
of maximal neutrino mass $m_\nu\simeq0.33\,$eV consistent with
large scale structure and CMB observations is only moderately
excluded.

The fact that a good part of the speculative scenarios of UHECR origin
are now ruled out makes the enigma of the highest energy particles
an even more exciting subject of study in our opinion.

\section*{Acknowledgments}
We acknowledge Sergey Troitsky and Michael Kachelrie\ss\  for
useful comments on a draft version of this paper.
We would like to thank Cyrille Barbot and Pasquale Blasi
for fruitful discussions. 
We also thank Douglas McKay, Teresa Montaruli and Cristian Spiering
for correspondence on the RICE, ANTARES and AMANDA-II experiments,
respectively.

This work was supported in part by the ACI ``Jeunes chercheurs''
entitled ``La Cosmologie: Un laboratoire pour la physique des
hautes \'energies'' awarded by the French research ministry.
The work of D.S. was supported by Deutsche Forschungsgemeinschaft
(DFG) within the Emmy Noether program. G.S. acknowledges hospitality
and financial research during his stay at the Max-Planck Institut
f\"ur Physik in Munich.

%%%%%%%%%%%%%%%%%%%%%%%%%%%%%%%%%%%%%%%%%%%%%%%%%%%%%%%
%%%%%%%%%%%%%%%%%%%%%%%%%%%%%%%%%%%%%%%%%%%%%%%%%%%%%%%%%%%%%%%%%%

\end{document}